# A NEW METHOD TO INDEX AND STORE SPATIO-TEMPORAL DATA

Guillermo de Bernardo, Departamento de Computación, Universidade da Coruña, A Coruña, Spain, gdebernardo@udc.es

Ramón Casares, Departamento de Computación, Universidade da Coruña, A Coruña, Spain, ramon.casares@udc.es

Adrían Gómez-Brandón, Departamento de Computación, Universidade da Coruña, A Coruña, Spain, adrian.gbrandon@udc.es

José R. Paramá, Departamento de Computación, Universidade da Coruña, A Coruña, Spain, jose.parama@udc.es

Abstract

*We propose a data structure that stores, in a compressed way, object trajectories, which at the same time, allow to efficiently response queries without the need to decompress the data. We use a data structure, called $K^2$-tree, to store the full position of all objects at regular time intervals. For storing the positions of objects between two time instants represented with $K^2$-trees, we only encode the relative movements. In order to save space, those relative moments are encoded with only one integer, instead of two (x,y)-coordinates. Moreover, the resulting integers are further compressed with a technique that allows us to manipulate those movements directly in compressed form. In this paper, we show an experimental evaluation of this structure, which shows important savings in space and good response times.*

# 1    INTRODUCTION

Nowadays, mobile phone and GPS devices are becoming more and more popular. Moreover, planes, ships, and even some cars also have devices informing about their position. Hence, there is a large set of information about the positions of moving objects (or individuals). However, this vast amount information represents an enormous challenge, because we need efficient ways to store and query these data.

For example, we can be interested in knowing "who was at the science museum at five o'clock?", or "who visited the cathedral between six o'clock and seven o'clock?". The spatio-temporal databases were developed to answer this type of queries. The two main types of queries issued in spatial databases are the *time-slice queries*, which return data at a specific time instant, and the *time-interval* (Pfoser, et al., 2000) queries, which return data between two time instants.

The spatio-temporal database field has been extensively studied (Šaltenis et al., 2000; Beckmann et al., 1990; Tao and Papadias, 2001), and thus they have reached a mature state in both academic and industrial level. In this work, we provide a new point of view in the field of spatial database field. We present a new spatio-temporal index for moving objects, which is able to answer queries very efficiently, but uses compact data structures to save space.

The classical setup of a classical database system is based on a set of data structures which uses the memory hierarchy starting at the disk. The key component to index, and thus efficiently query, is a tree-based structure, like for example, the Spatio-Temporal R-tree and the Trajectory-Bundle tree (Pfoser, et al., 2000) or the Time-Parameterized R-tree (Šaltenis et al., 2000).

Trees have several appealing characteristics. They provide logarithmic search times and a nice adaption to the memory hierarchy, since they are easy to split between main memory and disk. This last feature is a requirement in classical computing.

A new family of data structures called compact data structures aim at joining in a unique structure the data and the access methods. This structure is compact in the sense that the space consumption is kept as low as possible in order to be able to fit the entire structure in main memory (Jacobson, 1989). By fitting data and access methods in main memory, we make a better use of the memory hierarchy, since the upper levels have higher bandwidth and lower latency. For this sake, these data structures have to use complex mechanisms in order to compress the data, but at the same time, the compressed data must be queriable without a previous decompression, otherwise the data could not be fitted in main memory.

The main contribution of this work is the first (to the best of our knowledge) compact data structure to represent and query data storing moving objects. We will show that we can represent the data using only up to 25.59% of the space needed to represent them in plain form, at the same time that we can query them without decompressing the data with very good response times.

By providing a compact data structure to represent and query spatio-temporal information, we open a new way to deal with this type of information, aligned with the new tendencies in computer science designed to face the problem of processing and/or analysing vast amounts of information. The term Big Data is often used to refer this new way of computation, and one of its strategies is to perform an "in-memory" data processing, that is, to avoid costly disk transfers, by trying to fit all the information, or at least larger chunks of information, in main memory.

# 2    RELATED WORK

An object trajectory is often modelled as a sequence of locations of the object, each one represented as the (x,y) coordinates of the object in the space. Each of those locations corresponds to a time instant, and thus the trajectory is a vector of coordinate positions.

There has been much research work related to object trajectories. Most investigations were oriented to efficient access and index the data. For example, the 3DR-Tree (Vazirgiannis et al., 1998) is a spatiotemporal method which considers access time as another axis within the spatial coordinates. In this structure, a line segment represents an object at a location in a time interval. The 3DR-Tree is efficient processing interval queries, but inefficient processing instant queries.

The RT-Tree (Xu et al. 1990) is a structure where the temporal information is maintained at the nodes of a traditional R-tree. In this structure, the temporal information plays a secondary role because the query is directed by spatial information. Therefore, queries with temporary conditions are not efficiently processed (Nascimento et al., 1998).

The HR-Tree (Nascimento et al., 1999) uses an R-tree for each time instant, but trees are stored keeping in mind the overlapping between them. The main idea is that, given two trees, the most recent corresponds to an evolution of the oldest one and then the sub-trees that compose them can be shared between both trees. The best advantage of HR-Tree is that it can run instant time queries with a good performance, whereas this structure requires too much space.

The MV3R-Tree (Tao et al., 2001) is a structure based on the manipulation of multiple versions of R-Tree. This is an extension of the MVB-Tree (Becker et al., 1996).

The SEST-Index (Gutiérrez et al., 2005) is a structure that maintains snapshots of the state of objects for some instants of time along with event logs between these snapshots. The events provide more information than a simple change of position of the object. For example, it can indicate when an object enters or exits from a particular place or if an object crashes with another.

From the previous structures, the only one that has a relationship with our work is the last one. In this work, we also use snapshots and logs to represent object trajectories. The difference is that SEST-Index is a classical approach, whereas our structure is a compact structure designed to be fit in main memory.

## 3 COMPACT DATA STRUCTURES

As usual, compact data structures can be built on top of other data structures. We devote this section to describe the compact data structures that are the basic components of our proposal.

In order to understand the ideas explained in this paper, we need to know two essential operations over bit sequences (or bitmaps), which can be solved in constant time (Navarro and Mäkinen, 2007):
- $rank_b(B, p)$ counts the number of occurrences of the bit $b$ in the bitmap $B$ until position $p$. For example, $rank_1(T, 2)$ returns the number of ones that appear in the bitmap T up to position 2.
- $select_b(B, n)$ returns the position in the bitmap $B$ where is located the $n$-th occurrence of the bit $b$.

The $K^2$-tree (Brisaboa and Ladra, 2009) is a compact data structure to represent binary relations. In essence, it represents a binary matrix, as it can be in Figure 1. The $K^2$-tree structure takes advantage of several properties of a matrix: sparseness, large areas of zeros, and clustering of ones. It achieves very low space and supports efficient navigation over the compressed data.

The $K^2$-tree is conceptually a tree. To form the root node, the binary matrix is subdivided into $K^2$ submatrices of equal size. Each one of the $K^2$ submatrices is represented in the root node by one bit. The submatrices are ordered from left to right and from top to bottom. In Figure 1, we have four submatrices, for each submatrix, the root node has a bit. Each bit indicates if the submatrix has at least one 1 or not. Therefore, the root node of Figure 1 is 1011, since the first submatrix has 1s, the second one does not have 1s, and so on. The submatrices that have 1s are divided again, following the same method until the subdivision reaches the cell-level.

Instead of using a pointer-based representation, the $K^2$-tree is stored using two bit arrays:
- *T (tree)*: stores all the bits of the $K^2$-tree except those in the last level. The bits are placed following a levelwise traversal: first the $k^2$ binary values of the root node, then the values of the second level, and so on.

- *L (last level leaves)*: stores the last level of the tree. Thus, it represents the value of original cells of the binary matrix.

Another important data structure is the permutation. It is designed to efficiently answer two main operations $\pi$ and $\pi^{-1}$ over a sequence of numbers, where each number only appears one time throughout the sequence. The first operation allows us to obtain the number of a specified position in the sequence, whereas the second operation identifies the position in the sequence of a given value.

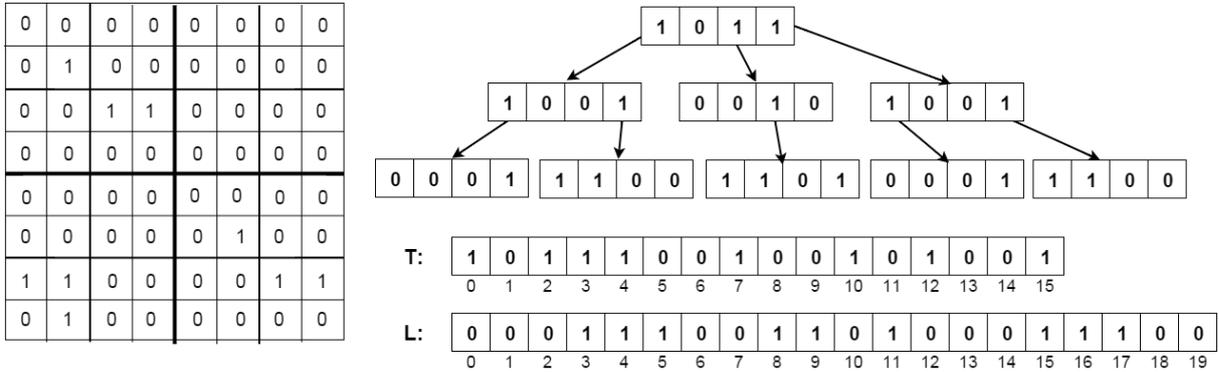

*Figure 1*  $K^2$-tree example where the K value is 2.

The permutation can be considered as an array of size *n*, where for each position or index we have a numeric value. This data structure allows us to return $\pi$ in constant time because we only need to get the value in the queried position. In order to answer $\pi^{-1}$, we do not need any additional structure. That is, the simple original sequence can answer $\pi^{-1}$. The procedure is based on a simple observation: if we take the number of a position as the index of the next position in a sequence of positions, any sequence will end up forming a cycle. In fact, in a permutation, there are usually several cycles.

For example, in Figure 2 we have the cycle (6, 9), since in position 6 it is stored a 9, and in position 9 we found a 6, closing the cycle. In that permutation, there are two more cycles (11, 12) and (1, 4, 3, 7, 8). The rest of elements are not in a cycle since they point to themselves.

If we want to know in which position is located the element with value 3, $\pi^{-1}(3)$, we access position 3, and we follow the cycle including that position until we find a position pointing to 3. That is, from position 3, we follow the cycle through 7, 8, 1, and finally 4, which is the position where the value 3 is found.

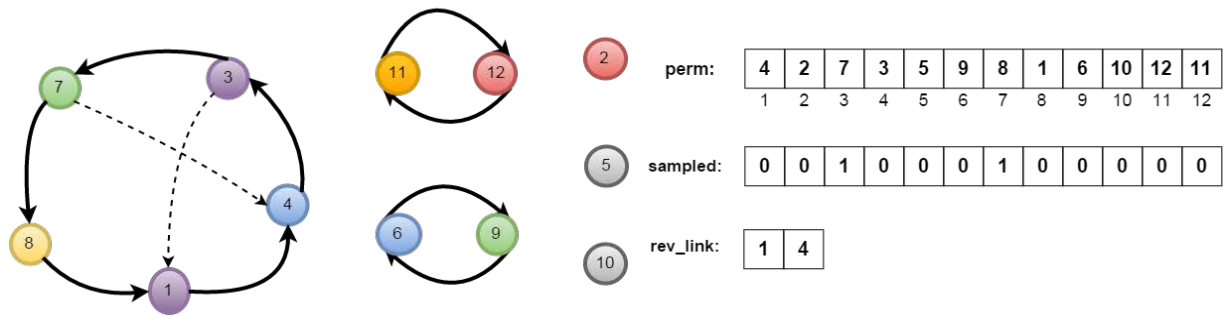

*Figure 2*  Permutation structure with links

A permutation could have a single cycle of length *n*. This would result in long searches to solve $\pi^{-1}$. In order to avoid this problem, the permutation is provided with two additional arrays (Munro, et. al., 2012). The first array is called *sampled* and it signals the starting positions of shortcuts that allow skipping some nodes of our cycle. The other array is called *rev_links*, it gives us the position where we jump when we find a shortcut in the cycle. If we use the above example, we do not need traverse the 7 and 8 states, we skip from state 3 to state 1, and then we reach the fourth position. Observe that to obtain $\pi^{-1}(3)$, we start the search at state 3, and we have to go through its cycle until it reaches 3 again, in fact, we are only interested in that last step, that is, when we close the cycle, so the shortcuts are designed in such a way that we are always sure that we cannot miss the last step of a cycle started in a given position. For example, if we want to obtain $\pi^{-1}(4)$, we start the search at 4, when we reach the state 3, we take the shortcut to state 1, just the state we are looking for, since it points to 4.

Finally, we present a statistical compression method called *(s,c)-Dense Code* (Brisaboa, et al., 2003). Statistical compression uses a *model* that gives the frequency of appearance of each source symbol (words, integers, etc.). Then a *compression scheme* assigns longer codewords to the less frequent source symbols and shorter codewords to the most frequent. Each original symbol is substituted by its codeword in the compressed file, which should have a prelude informing the correspondence between original symbols and codewords. The main feature of the *(s,c)-Dense Code* is that we can start decompression at any point, without decompressing from the beginning, a key feature in our method.

## 4 OUR PROPOSAL

In this proposal, we assume that objects can move freely over all the space, that is, they do not need to follow roads or any sort of net, although we can use it in that scenario. Examples of this could be boats navigating in the sea or birds flying on springtime migration.

We suppose that the time is discretized, in such a way that a time instant in our framework, would last a given period of real time. For example, for us, a time instant can last 5 seconds in terms of real time. The size of that period is a parameter. In a similar way, the two-dimensional space is divided into small cells (or tiles), where the size of each cell is another parameter.

First, we describe the data structures that hold the information, and then we describe the query algorithms.

### 4.1 Data structures

Our proposal uses two data structures called *snapshots* and *logs*. Snapshots give the full position of all objects at given time instant. Our method stores the full position of the objects at regular time intervals, whereas logs inform about the movements of the objects between two snapshots, as relative movements with respect to the last known position.

*4.1.1 Representation of snapshots*

A snapshot can store and index *m* objects that are located in the two-dimensional space in a compact way. Given that space is divided into cells, then it can be seen as a grid where each cell can hold objects. Observe that, if the cell size is sufficiently large, one or more objects can be located in one cell.

Each snapshot uses a $K^2$-tree to represent the distinct positions of our space. As explained, the $K^2$-tree is designed to represent a binary matrix. We map the binary matrix to the cells dividing the space, in such a way that each bit of the matrix corresponds to one of the cells of the space. If the bit is set to 1, this means that there is an object (or more) in the corresponding cell, whereas a 0, means that no object is present in the corresponding cell.

To represent a binary matrix, the only alternative compact data structure is a wavelet-tree (Brisaboa, et al., 2009). However, this structure has a poor performance when the number of cells set to 1 in a given column or row is greater than 1.

The basic $K^2$-tree only serves to check the presence or absence of objects in a given cell. In order to know which objects are in the cell, we need to add another structure. This new structure is formed by an array storing the objects identifiers (*perm*) and a bitmap called *quantity* or *Q*.

Observe that each 1 in the binary matrix corresponds, on the one side, to a bit set to 1 in the *L* bitmap of the K²-tree, and, on the other side, to one or more object identifiers. The list of object identifiers corresponding to each of those bits is stored in the *perm* array, where the objects identifiers are sorted following the order of appearance in *L*. That is, if L is 0101, and the first bit set to one represents a cell where there are two objects with identifiers 10 and 30, and the second bit set to one corresponds to a cell where we can find the objects 25, 28, and 29, then *perm* is 10, 30, 25, 28, 29. The *quantity* bitmap is aligned with *perm*, that is, for each element in *perm* we have a bit in *quantity*. This bit indicates with a 0 if the object in *perm* is the last object of a specific leaf, whereas a 1 signals that more objects exist. Therefore, *quantity* for our previous case is 10110. The first 0 corresponds to the object 30, which is the last object identifier corresponding to the first cell, whereas the second 0, corresponds to 29, the last object identifier of the objects present in the second cell.

With this information we can answer two queries:
- Search the objects of the *n*-th leaf: we count the number of 1s in the array of leaves *L* until the position *n*, this gives us the number of leaves with objects up to the *n*-th leaf, $x = rank_1(L, n)$. Then we calculate the position of the *(x-1)*-th 0 in *Q*, which indicates the last bit of the previous leaf, and we add 1 to get the first position of our leaf, $p = select_0(Q, x-1)+1$. *p* is the position in *perm* of the first object identifier corresponding to the searched position. From *p*, we read all the object identifiers aligned with 1s in *Q*, until we reach a 0, which signals the last object identifier of that leaf.
- Search the leaf in *L* corresponding to the *k-th* position in *Q*: First, we calculate the number of leaves before the object corresponding to *k*, that is, we need to count the number of 0s until the position before *k*, $y = rank_0(Q, k-1)$. Then we search in *L* the position of the *y+1-th* 1, that is, $select_1(L, y+1)$.

The process to get the objects of a cell is very efficient; we only require to find the leaf in the K²-tree ($O(h)$ where h is the height of the tree, that is, $O(\log n)$) and search the objects of this leaf, using *perm* and *Q*.

However, when we want to know the position of an object, with the structures shown so far, we need to search sequentially in *perm* the position of the searched object. To avoid this sequential search, we add a permutation over *perm*. The permutation allows us to find an object in the array ($\pi^{-1}$), spending less time than a sequential search, while keeps the constant time required to retrieve the object associated with a cell ($\pi$).

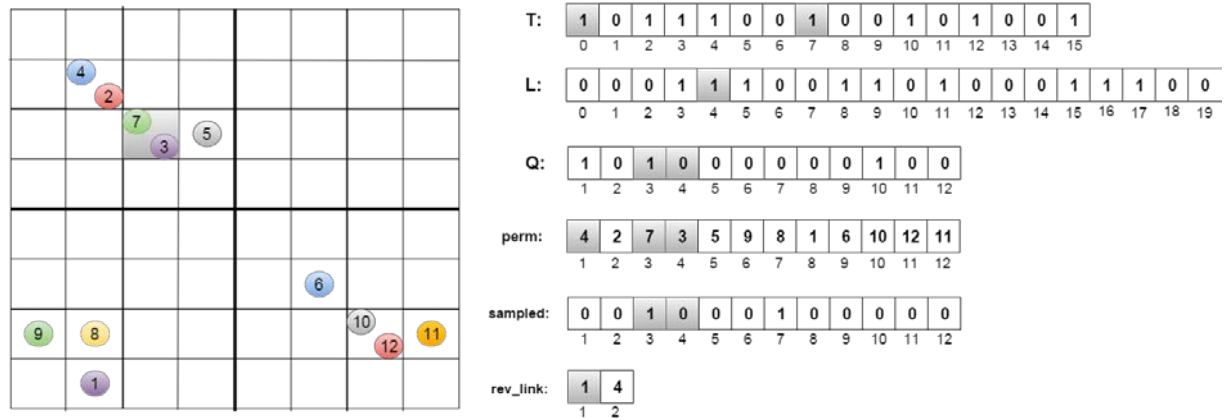

Figure 3. Snapshot example with the K2-tree and the permutation.

For example, assume that we have the binary matrix and the associated K²-tree of Figure 1. Figure 3 shows an example of object identifiers matching that matrix. In addition to the arrays *T* and *L*, we add the array *Q* and the array of object identifiers *perm*. Finally, over *perm*, we construct our permutation with its cycles and over that cycles, we add three links, in order to improve the performance of the queries. In this example, we are using the permutation shown in Figure 2.

If we want to find the objects that are in the cell corresponding to the third row and third column (3,3), we start the search at the root of the K²-tree, and we traverse it downwards until we reach the leaf representing that cell. First, we compute which child of the root node overlaps the cell (3,3), in this case, it is the first child, that is, position 0 of *T*. That position is set to 1, so we continue the search in the next level. In the second level, we check which submatrix overlaps the cell (3,3), and we obtain the fourth child[a1], which is represented by the bit at position 7 of *T*. Since that bit is set to 1, we continue, reaching the last level. Finally, the leaf corresponding to the cell (3,3) is the first child [a2]of the considered submatrix (position 4 in *L*) and its value is a 1, hence the cell holds objects.

Next, we need to know which positions of *Q* are linked with that leaf. In order to search these positions we count the number of leaves with objects before our leaf, $rank_1(L,p-1)$ where *p* is the position of our leaf in *L*. In our case, *p* is 4 so we obtain $rank_1(L,3)=1$, therefore, we know that there is one leaf with objects before our leaf. Then, we find in *Q* where the previous leaf ends, that is $select_0(Q, k)$, *k* is the number of leaves before the leaf we are searching for. Thus, we calculate $select_0(Q, 1)=1$. We know that next position is the start of our leaf, so we sum one and we obtain a 2. Finally, we access that position in the permutation, *π(2)* (the object with id 7) is the first object of the leaf and, therefore, the first object that is located in the cell. We continue obtaining the rest of objects sequentially until we obtain that the aligned value of the *Q* bitmap is a 0. In our example, the next bit is *Q[3]*, and it already stores a 0 so *π(3)* will be the last element of the cell. *π(3)* returns a 3, so the objects 7 and 3 are in the cell (3, 3).

Although the K²-tree was designed to traverse the structure from top to bottom, with a small modification, we can go from bottom to top. In this way, we can obtain the position in the grid of a given object.

For example, we can search for the cell where is placed the object 3. For this sake, we access to the permutation and execute the operation $π^{-1}(3)$. We start the search at position 3, and we follow the cycle. We observe that we have a link between 3 and 1 because in the sampled array we have a 1, and in *rev_links* the value is 1, which indicates that the link points to the state 1, so we skip the 7 and 8 and we go to the state 1. We need to continue because *π(1)=4*, hence we advance to next state. In *π(4)*, we find the value 3, the value which we are searching for. Therefore, the position of object 3 is the fourth position in *perm*.

Now, we search the corresponding leaf of the object. We calculate the number of leaves previous to the object, thus, we need to count the number of 0s until position 3 in *Q*, $rank_0(Q, 3)=2$. Then we search in *L* the position where the leaf is, hence we calculate $select_1(L, 3)= 4$, so *L[4]* is the leaf where is the object 3. Then, we traverse the tree from the bottom to the top and we obtain the path that indicates that it is in the cell (3,3).

### 4.1.2   Representation of the log of relative positions

Although the K²-tree allows to represent the position of all objects in a compact way, it would be prohibitive, in space terms, to use a snapshot (a K²-tree and the associated *perm* and *Q* data structures) for each time instant. Thus, in order to avoid this problem, we use a parameter to establish the number of time instants between two snapshots and, in order to keep the information about the movements of objects between two snapshots, we add a light structure called *log*.

The log stores the relative movements of all objects for all time instants between two snapshots. The objects can change their positions in the two Cartesian axes, so we would reserve two integers for every movement in the log, one for x-axis and another for y-axis. However, in order to save space, we encode the two values with a unique positive integer.

For this sake, we enumerate the cells around the actual position of an object, following a spiral where the origin is the initial object position, as it is shown in Figure 4.

| 42 | 43 | 44 | 45 | 46 | 47 | 48 |
|----|----|----|----|----|----|----|
| 41 | 20 | 21 | 22 | 23 | 24 | 25 |
| 40 | 19 | 6  | 7  | 8  | 9  | 26 |
| 39 | 18 | 5  | 0  | 1  | 10 | 27 |
| 38 | 17 | 4  | 3  | 2  | 11 | 28 |
| 37 | 16 | 15 | 14 | 13 | 12 | 29 |
| 36 | 35 | 34 | 33 | 32 | 31 | 30 |

|   |   |   |   |   |   |
|---|---|---|---|---|---|
|   |   |   |   |   | $5^2$ |
|   |   |   |   | $3^2$ |   |
|   |   |   | 0 | $1^2$ |   |
|   |   | $2^2$ |   |   |   |
|   | $4^2$ |   |   |   |   |
| $6^2$ |   |   |   |   |   |

*Figure 4    Spiral matrix to encode relative positions.*

For example, we have an object that moves between instant $t_0$ and $t_4$. The $t_0$ corresponds with a snapshot so we store its absolute position in the snapshot, but the following instants should be stored in the log. Assuming that we have the following (x,y) coordinates:

$$\{ t_0: (5,6) ; t_1: (7,8) ; t_2: (5,8) ; t_3: (3,10)\}$$

we calculate the movements between the time instants:

$$\{t_1-t_0= (2,2); t_2-t_1=(-2,0); t_3-t_2=(-2,2)\}$$

Then, we encode these shifts with the spiral $\{t_1-t_0= 24; t_2-t_1=18; t_3-t_2=16\}$, which are the values stored in the log.

Observe that the basic idea is that with the spiral, we save space by using only one integer, instead of the two integers needed when using pairs of coordinates of the Euclidean space. A problem would arise if the displacement of the object between two consecutive time instants is large. In that case, the value obtained using the spiral encoding will be a large number, which might require a considerable number of bits to represent it. However, it is likely that the displacement between two time instants will be small, and thus, in most cases, a short integer will be enough. In addition, [a3]instead of using 32 bits (or 64 bits) for each relative movement, we exploit the repetitiveness in the values of movements to obtain additional compression, since some movements are more frequent than others. For example, in the Chilean coast is more frequent that a boat navigates towards the North than towards the East. Thus, we do not store the relative movements using plain integers, instead that sequence of integers is compressed using the (s,c)-Dense Code technique.

Recall that one of the main features of this compression technique is that we can start the decompression at any given point. This is needed to be able to decompress the log starting just after a snapshot, avoiding to start the decompression from the beginning.

*4.1.3    Disappearance and reappearance*

In the real world, sometimes objects stop emitting signals or send erroneous information about their location. The last problem is easy to avoid because we can process and clean the dataset removing these incorrect data. However the first problem forces us to add two new possible movements:
- Relative reappearance: it occurs when a missing object sends a new position, but we can obtain its new position with a relative movement with respect to the last snapshot or with respect to another log position. We only use relative reappearance with objects that disappear and reappear between the same snapshots. For example, if we have snapshots every two hours and we have a snapshot at eight o'clock and another at ten o'clock, and an object disappears at eight o'clock and reappears at nine o'clock, we can add a relative reappearance. However, if it reappears at eleven o'clock, we cannot add a relative reappearance.
  It is important to notice that we can have more than one relative reappearance per object, so we need to add a new array to the snapshot structure. This array saves for each object the time elapsed and the relative position between the instant when the object disappears and the instant when the object reappears. Every relative reappearance in the log points to its related position in this array, allowing us to recover all the information about the reappearance.
- Absolute reappearance: it occurs when a missing object sends a new position and it had disappeared before the last snapshot. In this case, we cannot know where the object is located because we have not any reference. We could find the last snapshot which contains the object, calculate the log movements until its disappearance and then calculate the new position decoding the spiral identifier, however it would be very expensive. Therefore, we store the absolute position of the object and the instant when the object reappears without any compression, it should be noted that an object can only have one absolute reappearance between snapshots, so we only have to store these three values, at most, once per object.

## 4.2 Querying

We can distinguish three types of queries: search where is an object in a given time instant, search the objects that are in a zone at a time instant (*time slice*), or in a time interval (*time interval*).

### 4.2.1 Search an object in a given time instant

First, we obtain the immediate previous snapshot to the query time instant. Then, we need to found in the array of leaves which leaf correspond to the object. We do that using the permutation, as explained before. Once we find the leaf linked to the object, we through bottom-up the $K^2$-tree and, in this way, we retrieve the position of the object at the snapshot instant. Then we apply the movements of the log over this position up to the query time instant.

It is important to notice that if the time instant to check belongs to a snapshot we do not need to access the log. In addition, our structure can answer efficiently queries about the trajectory of an object between two instants.

### 4.2.2 Time slice query

The aim of this query is to know which objects are at a time instant $t_q$ in a region of space. We can distinguish two cases. First, if the time instant belongs to a snapshot, we only need to traverse the $K^2$-tree until the leaves, visiting those nodes that intersect with the checked area. When we reach the leaves we can retrieve from the permutation the objects that are in this area.

The second case occurs when $t_q$ is between two snapshots $s_i$ and $s_{i+1}$. In this case, we check in $s_i$ a region larger $r'$ than the query region $r$. This region $r'$ is built keeping in mind that all the objects cannot move faster than the fastest object of the dataset. Thus, we limit our region $r'$ to the space where an object has chances to reach the region $r$ at the time instant $t_q$. Then we get the objects which belong to the region $r'$ in the snapshot $s_i$, these are the candidates to be in the region $r$ at $t_q$. We follow the movements of the objects with the *log* and when we observe that an object cannot reach the final region, we discard it. Hence, we limit the number of objects to follow and we do not waste time with those objects that cannot be in the region $r$ at instant $t_q$.

For example, we suppose that we want to find the objects that are located in a region $r$ at time instant $t_q$. Suppose that the closest previous snapshot to $t_q$ is at time instant $t_0$. Assume that the fastest object can move only to one adjacent cell, in the period of time between $t_0$ and $t_q$. Therefore, we expand our query area in $t_0$ to its adjacent cells, as we can see in Figure 5. In this region, we have three elements (1,2,3) which we call *candidates*, so we discard the rest of objects (4, 5), because they do not have any chance to be in the region $r$ at time instant $t_q$. Then, we control the *candidates* with the log until reach the time instant $t_q$, where we observe that the object with id 3 is outside the region $r$ and the rest of candidates are the result of the query.

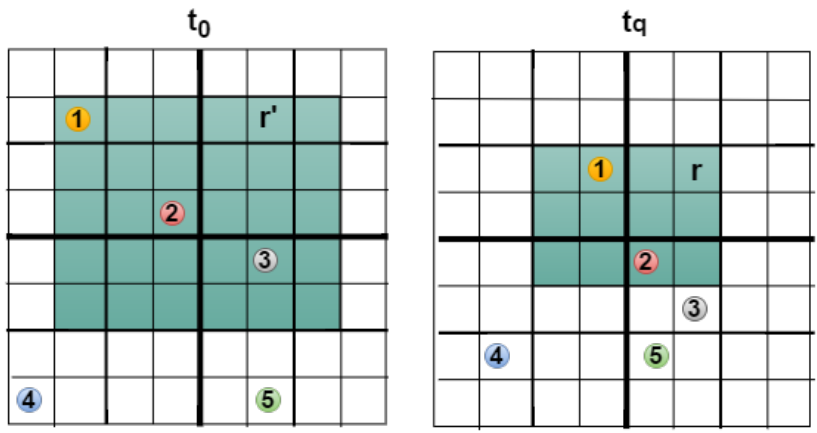

*Figure 5    Example of expanded region r' and query region r in a time-slice query.*

*4.2.3    Time interval query*

This query returns the objects in a region *r*, in a time interval *[t₀, tₙ]*. To solve it, we divide the time interval into smaller intervals. With each subinterval, we run a subquery. The first subinterval will be the interval *[t₀, t_{s1}-1]* where $t_{s1}$ is the first snapshot after $t_0$, the second interval will be *[t_{s1}, t_{s2}-1]* where $t_{s2}$ is the second snapshot after $t_0$, and so on. When we run these subqueries, we obtain the objects which are in the region *r* during the subinterval, and then they are part of the final result, so we do not need to check their position in the next subqueries. Once we obtain all subquery results, we join the partial results to obtain the final result. It is important to notice that this query is very similar to time slice, its difference is that we need to consider the larger region *r'* for every snapshot of the subqueries except the last one. Although this query is more expensive than the time slice query, it has a good performance because we take advantage of rejected objects in each subquery, which are no longer considered.

## 5    EXPERIMENTAL EVALUATION

In this work, we pursue to develop a proof of concept of our structure. Observe that there is not a clear competitor since our approach aims at storing the complete collection in main memory and thus, the saving of space is one of our main targets. Classical approaches to the same problem are disk based, and hence, they are not worried about space consumption. Therefore, the goal of this experimental evaluation is to show the compression level and, at the same time, that we are capable of querying the compressed data.

We used a real dataset, which we obtained from the site http://chorochronos.datastories.org/?q=node/81. This a real dataset that provides the location signals of 4,856 boats moving during one month in the region of the Greek coast. As any basic compression technique, our method requires the presence of repetitive patterns, which are the source of compression. In this dataset, ships follow many times the same course, and thus our compression method can take advantage of this fact.

Observe that our structure deals with the positions of objects at regular intervals, however in the dataset, boat signals are emitted with distinct frequencies or, moreover, they can send erroneous information. Therefore, we need to clean the dataset and fill it with positions that we can interpolate from previous/posterior signals. For example, when a boat sends a position after a large period of time without sending its position, and the new position coincides with the last received position, we can assume that the boat stood in the same position all the time, hence, we fill the dataset with the position received after the period without signals. Another case happens when a boat sends an erroneous location. We can detect it, if the boat moves faster than 60 knots, which is the fastest speed of our dataset. Therefore, we can discard this location and interpolate the real position with previous and next positions.

Every position emitted by a boat is discretized into a matrix where the cell size is 30x30 meters. This discretization is required, since the K$^2$-trees that store the snapshots can represent binary rasters, and any raster model represents the space as a tessellation of squares, each representing a square of the real space with a given size, in our case 30x30 meters.

In addition, we normalize the time of signals emitted using intervals of 1 minute. This another requirement of our method, since we represent the position of the objects at regular time instants. Therefore, we assign for each boat the nearest position every 1 minute with a margin before and after of 30 seconds. For example, if a boat sends information between the second 30 and the second 90 we assign to the time instant 1 the nearest location to the second 60. Whereas if it does not emit between 30s and 60s, we do not assign any position to this boat at time instant 1. With this data normalization, we obtain a matrix with 920,078,376 cells, 33,544 in the x-axis and 27,429 in the y-axis, and the data corresponds to 44,640 instants of time.

With these data, which occupy 515.89MBytes, we run two kinds of experiments using a program in C++ in a 1.70GHzx4 Intel Core i5 computer with 8GBytes of RAM and an operating system Linux of 64bits. In the first experiment, we run a set of queries over a structure built with different snapshot frequencies. These queries are composed by 100 random searches of an object at a time instant and 50 random time slice queries. We calculate the size of the structure (MBytes), the compression ratio and the response times (milliseconds). The compression ratio value indicates the percentage of space occupied by the compressed version with respect to the original dataset. For example, with a snapshot period of 120, our structure occupies 25.59% of the spaced consumed by the original dataset, whereas when the period of snapshot is 15, our structure occupies a 143%, that is, it occupies even more than the original dataset, specifically 43% more.

| Snapshot period | Structure size (MB) | Compression ratio | Average time of object search (ms) | Average time of time slice query (ms) |
|---|---|---|---|---|
| 120 | 132.03 | 25.59% | 0.01784 | 5.29740 |
| 60 | 218.59 | 42.37% | 0.01422 | 2.93332 |
| 30 | 392.35 | 76.05% | 0.01347 | 2.51360 |
| 15 | 739.96 | 143.43% | 0.01329 | 2.20784 |

*Table 1     Compression ratio and response time (ms).*

As it is shown in Table 1, in the first experiment we can observe as the compression ratio improves when we increment the snapshot period, however, we penalize the response time.

|  | Average time of object search (ms) | Average time of time slice query (ms) |
|---|---|---|
| Result in the snapshot | 0.01262 | 0.45002 |
| Result in the first quarter of log | 0.01508 | 3.59028 |
| Result in the middle of log | 0.01760 | 3.72642 |
| Result in the third quarter of log | 0.02010 | 5.52748 |

*Table 2     Response time in different situations.*

In the second experiment, we want to show the behaviour of our structure when we solve queries using only the snapshot or when accessing the log as well. In this experiment, we used the setup of the best compression ratio of the first experiment, that is, the structure with a snapshot period of 120. Over this structure, we run queries which we can be solved directly in a snapshot and others that require traversing up to the first quarter, the middle, and the third quarter of the log. For each case, we run 50 random queries, where we obtain at least one result. Then, we measure the average time in milliseconds of searches for an object and time slice queries. Table 2 shows the results.

As it is shown in Table 2 we can observe that the time of both kinds of queries is better when we are searching in a snapshot than when we need to access the log. In the case of time slice query, we can see that the response time grows faster than in object searches. It is due to the tracking of objects that can reach the queried region at the considered time slice, whereas in searches by object, we only must follow a given object.

Finally, observe that all response times are very appealing as they are below 6 ms., less than just one disk access, which would require around 10-20 ms.

# 6    IMPROVEMENTS

In order to improve the performance of our structure, we designed two enhancements that will allow us to process the queries more efficiently.

## 6.1    Bidirectional log

Consider an inquiry that queries a time instant $t_k$, which is between the snapshots $s_i$ and $s_{i+1}$. Let us suppose that $t_k$ is closer to $s_{i+1}$ than to $s_i$, in that case, we must read a large portion of the log, since we have to start from $s_i$ up to the time instant that we are checking. However, if would desirable to have the chance to start the search at $s_{i+1}$ and traverse backwards the log until we reach $t_k$. In this way, we could reduce the searches over the log to, at most, half of the size of the log between two snapshots.

In order to allow those backwards searches, we simply add, to each portion of the log between two snapshots $s_i$ and $s_{i+1}$, the full position of all objects in the time instant immediately before the time instant corresponding to $s_{i+1}$. With this, we can perform a backwards navigation, for example, if we have the snapshots $s_0$ and $s_1$ delimiting the time instants between $t_0$ and $t_8$, and the objects perform these movements:

$\{t_1-t_0= (2,2); t_2-t_1=(-2,0); t_3-t_2=(-2,2); t_4-t_3=(-1,2); t_5-t_4=(-2,2); t_6-t_5=(-2,2); t_7-t_6=(-1,1); t_8-t_7=(0,2)\}$

Assuming that the position previous to the time instant of $s_1$ is *(-8, 13),* and we want to know where is the object at the time instant $t_6$, since we know that the position in $t_8$ is *(-8, 13)* and we have the shift between $t_8$-$t_7$, we can revert the changes from the last snapshot until the time instant $t_6$. Hence, if we run: *(-8,13)-(0,2)-(-1,1)= (-7, 10),* the position of the object at the time instant $t_6$.

Now, we have to solve the problem of disappearances. If we find an identifier that indicates an object reappearance, we simply follow the pointer to the reappearance array. In this array, we retrieve the relative position and the time between the disappearance time instant and the reappearance. Thus, we never lost the synchronization with the time.

In addition, this structure can help us when we run time-slice queries because we can use the closest snapshot which allows us to reduce the region area $r'$ and the number of objects to follow during the query execution.

## 6.2    Accumulators

To reduce the running times of queries, we can add accumulators at regular time intervals in the logs. An accumulator sums all the logs values between a snapshot and the time instant where the new accumulator is added. Hence, now instead of iterating over the log from the snapshot in the left extreme of a log portion, and summing the object positions until a selected instant, we obtain the accumulator before the checked instant and we sum the relative positions from the accumulator to the considered time instant. In fact, if the right snapshot is closer, this process can be done in the same way from that snapshot, but subtracting.

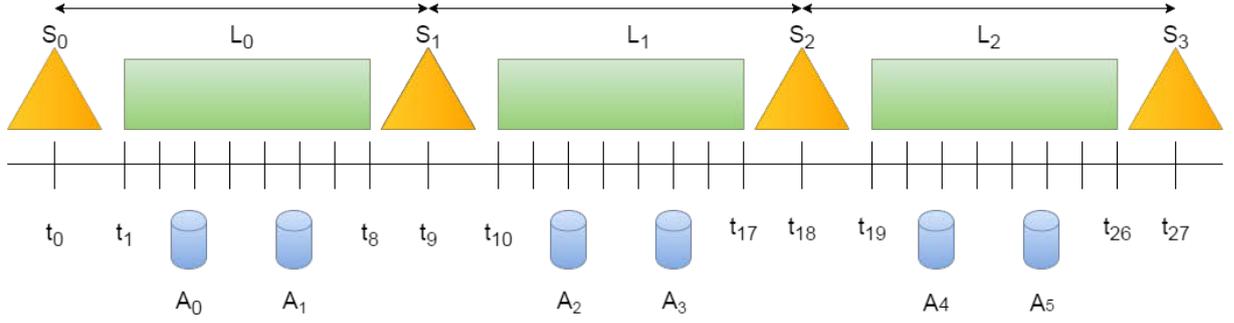

*Figure 6.     Example of accumulators $A_x$ in the logs $L_x$ between snapshots $s_i$ and $s_{i+1}$*

For example, in Figure 6, assume that we want to know the position of a boat at time instant $t_{13}$. First, we need to find the object position in the nearest snapshot $s_1$. Next, we obtain the value in the accumulator $A_2$. Finally, after adding the values obtained in that structure, we add the relative value obtained from the instant $t_{13}$.

# 7     CONCLUSIONS

We have developed a new data structure to represent the object trajectories. This proposal is based on the compact data structures approach. That is, we aim at keeping in main memory the whole structure and the access methods in order to avoid the disk access. By fitting data and access methods in main memory, we make a better use of the memory hierarchy, since the upper levels have higher bandwidth and lower latency. If we decompress it completely, we will be forced save parts in disk, since it is likely that the uncompressed structure will not fit completely in main memory. However, with our approach, we can only decompress the structure parts that we need to answer the query which we are running.

This structure obtains compression ratios up to 25.59%, hence we reduce substantially the space consumption, having more chances to fit the data structure in main memory. In addition, we obtain a good performance searching the object positions at a time instant and answering time slice queries. During the experimental evaluation, we can observe that if we reduce the snapshot period the structure size grows and the queries are faster. This is caused by the average distance of the log that we need to traverse in order to response the queries.

As a future work, it would be very interesting the development of the bidirectional log and the accumulators structure. Thus, we would solve queries that access and traverse the log with a better performance.

## Acknowledgments

This work was funded in part by European Union's Horizon 2020 research and innovation programme under the Marie Sk lodowska-Curie grant agreement No 690941, Ministerio de  Economía y Competitividad under grants [TIN2013-46238-C4-3-R], [CDTI IDI-20141259], [CDTI ITC-20151247], and [CDTI ITC-20151305] and Xunta de Galicia (co-founded with FEDER)   under grant [GRC2013/053].